\documentclass[prb,twocolumn,superscriptaddress,showpacs]{revtex4}

\usepackage{graphicx}
\usepackage{dcolumn}
\usepackage{amsmath}
\usepackage{color}

\begin{document}
\title{ESR evidence for partial melting of the orbital order in LaMnO$_3$ below the Jahn-Teller transition}

\author{S.~Schaile}
\author{H.-A.~Krug von Nidda}
\author{J.~Deisenhofer}
\affiliation{Experimentalphysik V, Center for Electronic
Correlations and Magnetism, Institute for Physics, Augsburg
University, D-86135 Augsburg, Germany}
\author{M.~V.~Eremin}
\affiliation{Institute for Physics, Kazan (Volga region) Federal University, 420008 Kazan, Russia}
\author{Y.~Tokura}
\affiliation{Quantum-Phase Electronics Center and Department of Applied Physics, University of Tokyo, Tokyo 113-8656, Japan}
\affiliation{RIKEN Center for Emergent Matter Science (CEMS), Wako 351-0198, Japan}
\author{A.~Loidl}
\affiliation{Experimentalphysik V, Center for Electronic
Correlations and Magnetism, Institute for Physics, Augsburg
University, D-86135 Augsburg, Germany}

\date{\today}

\begin{abstract}
We report on high-temperature electron spin resonance studies of a detwinned LaMnO$_3$ single crystal across the Jahn-Teller transition at $T_{\rm JT}$ = 750 K. The anisotropy of the linewidth and g-factor reflects the local Jahn-Teller distortions in the orbitally ordered phase. A clear jump in the linewidth accompanies the Jahn-Teller transition at $T_{\rm JT}$ = 750 K confirming that the transition is of first order. Already at $T^*$ = 550 K a significant decrease of the reduced linewidth is observed. This temperature scale is discussed with respect to the interaction of the $e_g$-electrons of the Mn$^{3+}$-ions and the elastic field of the cooperative distortions. Our results support a partial melting of the orbital order along the antiferromagnetically coupled $b$-axis at $T^*$. The remaining two-dimensional orbital ordering within the ferromagnetically coupled $ac$-plane finally disappears together with the cooperative distortion at $T_{\rm JT}$. Moreover in our discussion we show that elastic strain field interactions can explain the melting of the orbital order and, thus, has to be taken into account to explain the orbital ordering in LaMnO$_3$.

\end{abstract}



\maketitle
\section{Introduction}
The antiferromagnetic Mott-insulator LaMnO$_3$ is famous for being the mother compound\cite{Goodenough55,Woll55} of a large family of doped perovskites ($R_{1-x} A_x$MnO$_3$ with $R$ = trivalent rare-earth ion, $A$ = divalent alkali earth Ca, Sr, Ba) showing a variety of interesting physical properties like colossal magnetoresistance,\cite{Kusters1989,Helmolt1993} charge/orbital ordering,\cite{Goodenough1961,Murakami1998,Yamada1996} and multiferroicity.\cite{Kimura2003,Noda2006,Hemberger2007}
These interesting physical properties arise from the complex interplay of spin, orbital, charge and structure degrees of freedom which is triggered by the Jahn-Teller (JT) active Mn$^{3+}$ ions in LaMnO$_3$. The fivefold degenerate $d$-levels of Mn$^{3+}$ are split by the cubic crystal field in octahedral oxygen coordination into a lower $t_{2g}$-triplet, half occupied by three electrons, and a higher $e_g$-doublet, occupied by a single electron only. The residual degeneracy of the $e_g$-doublet is lifted by the cooperative Jahn-Teller distortion of the MnO$_6$ octahedra within the orthorhombic ($Pnma$) structure.\cite{Goodenough55} The corresponding orbital order accounts for an $A$-type antiferromagnetism with a N\'{e}el temperature $T_{\rm N} = 140$~K. The cooperative Jahn-Teller derived orbital order survives up to $T_{\rm JT} =750$~K. At this temperature LaMnO$_3$ undergoes a structural transition from the orbitally ordered $O'$ to the orbitally disordered $O$ phase which is nearly cubic.\cite{Rodriguez98}
The transition from the $O'$ phase to the $O$ phase is accompanied by a volume collapse of 0.36\%\cite{Chatterji03} which does not change the Mn-O bond length.\cite{Souza04}

The volume collapse, together with step-like anomalies in other experimental quantities like electrical resistivity and magnetic susceptibility\cite{Zhou99} indicates a first-order character of the JT transition. This was corroborated by the theoretical treatment of the anharmonic coupling between the pseudospin of the Mn$^{3+}$ $e_g$ states, the staggered JT distortion, and the volume strain coordinate.\cite{Maitra2004} Moreover it was shown that the first-order transition transforms into a second-order one dependent on the model parameters and was experimentally realized by doping experiments.
This finding was supported by results obtained from different experimental techniques on La$_{0.95}$Sr$_{0.05}$MnO$_3$ single crystals, a series of Sr-doped powder samples and LaMnO$_3$ thin films.\cite{Deisenhofer03,Mitchell96,Song02}
Besides this general assessment concerning the character of the JT transition, the dynamical properties observed in a rather broad temperature range around $T_{\rm JT}$ deserve a closer inspection.
Recent Nuclear Magnetic Resonance (NMR) experiments\cite{Trokiner13} show changes in the spin-relaxation and the Knight-shift of the ${}^{17}$O ligand ions already far below $T_{\rm JT}$, which are interpreted as a melting of the orbital order. 
But as the observed magnetic shift at the ${}^{17}$O-site is related to the so called "`short range effect"' the upturn of the echo decay rate may also be a signature of the disapperence of short range spin order. Moreover, a peak in the temperature dependence of transverse relaxation rate at the oxygen sites (${}^{17} T_{2}^{-1} $) at $T\cong 900$~K indicates collective excitations above $T_{\rm JT}$, matching non-zero local distortions found by neutron diffraction and x-ray absorption measurements.\cite{Wdowik11,Sanchez03} In contrast to the NMR and studies of the crystal structure, the ESR linewidth is directly determined by the on-site crystal field parameters D and E and, thus, is not related to short range spin-order effects. It probes directly the orbital state at the Mn-site and provides direct information on the orbital ordering.

With a new high sensitivity, high temperature setup\cite{Schaile12} we are now able to present high-temperature electron spin resonance (ESR) measurements up to 1000~K taken in a detwinned LaMnO$_3$ single crystal. The anisotropy of ESR resonance field and linewidth observed in the orbitally ordered $O^{\prime}$ phase is in good agreement with the results obtained earlier in lightly Sr doped crystals.
The high-temperature data bear further important information on the orbital degrees of freedom close to the structural transition.
Besides a rather abrupt breakdown of the anisotropy on crossing $T_{\rm JT} = 750$~K into the $O$ phase, we find evidence for a change of the anisotropic interactions already above $T^*=550$~K indicating the onset of strong orbital fluctuations at temperatures far below the structural transition at $T_{\rm JT}$. Usually two types of interactions are considered to be responsible for the transition at $T\approx T_{JT} $: orbital dependent superexchange and the Jahn-Teller coupling with concomitant local distortions. However, it was pointed by Sanchez \textit{et al.}\cite{Sanchez03} that these two mechanisms are not enough in the description of that unusual phase transition. In this study we consider, therefore, the coupling of the orbital degrees of freedom to the elastic strain field.

\section{Experimental details}

LaMnO$_3$ single crystals have been grown by the floating-zone method in a halogen-lamp image furnace. A twinned single crystal was detwinned as described in Ref.~\onlinecite{Tokura01}. For single-crystal measurements the crystal was cut into thin plates alongside the crystal axes determined by Laue patterns. Measurements of polycrystalline LaMnO$_3$ were done on crushed single crystals. The magnetic susceptibility of the samples was measured using a superconducting
quantum interference device (SQUID; Quantum Design MPMS5) at temperatures $2 \leq T \leq 800$~K and in external magnetic fields up to 50~kOe.

ESR detects the power $P$ absorbed by the sample from the transverse magnetic
microwave field as function of the static magnetic field $H$. The
signal-to-noise ratio of the spectra is improved by recording the
derivative $dP/dH$ using lock-in technique with field modulation.
For ESR measurements in the temperature region $300 \leq T \leq 1000$~K a Bruker Elexsys II spectrometer equipped with a Bruker ER 4114 HT X-Band ($\nu = 9.3$~GHz) cavity was used. For measurements in the temperature region $4 \leq T \leq 300$~K a Bruker Elexsys I spectrometer alternatively equipped with a Bruker ER 4102 ST X-Band ($\nu = 9.3$~GHz) cavity and a Bruker ER 5106 QT Q-Band ($\nu = 34$~GHz) cavity was used together with matching Oxford continuous He gas-flow cryostats.

Due to large temperature gradients at high temperatures the sample temperature is
corrected by calibration measurements at the actual sample site in
the high-temperature cavity. To check the influence of different
environmental conditions and temperature on the oxygen content of
the samples at high temperatures, measurements have been performed
in pure Ar at 0.7 bar pressure adjusted at room temperature and in air
(open quartz tube) for polycrystalline samples.
Possible changes in the oxygen content of the samples during the measurements at elevated temperatures are excluded by returning to lower temperatures during the measurements. Changes of the oxygen content would strongly affect the magnetic properties of the samples\cite{Prado99} and thus, result in perceivable steps in the linewidth. It must be noticed that the used high-temperature cavity is not capable for the low cooling rates necessary for keeping the detwinned state by cooling across the Jahn-Teller (JT) transition at 750~K.

\section{Experimental Results}

\begin{figure}[b]
\includegraphics[width=0.5\textwidth]{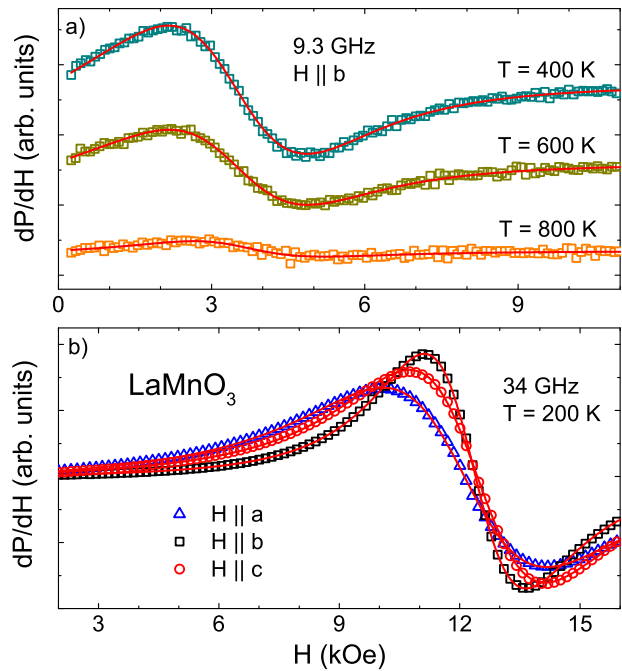} \caption{\label{fig:spek} (Color online) ESR-spectra of LaMnO$_3$. The solid lines refer to the corresponding fit curves using a Dyson line shape. a) X-Band ESR-spectra at different temperatures below and above $T_{\rm JT} =750 $~K. The magnetic field is aligned parallel to the $b$-axis of the crystal.
b) Q-Band ESR-spectra at $T=200$~K with the magnetic field aligned parallel to the crystallographic axes.}
\end{figure}

From the large number of recorded spectra a selection depicted in Fig.~\ref{fig:spek} illustrates the temperature dependence across the Jahn-Teller transition and the anisotropy in the orbitally ordered phase in LaMnO$_3$. Note that due to low Q-factor cavities used for the high-temperature measurements, the signal-to-noise ratio is rather low compared to ESR measurements in standard ESR cavities. Moreover, the skin effect arising from the conductivity of the single crystals reduces the penetration depth of the microwave radiation and, thus, weakens the recordable signal in comparison to measurements in polycrystalline samples -- especially above $T_{\rm JT}$, where the conductivity significantly rises.\cite{Souza07}
For all temperatures, orientations, and frequencies, the ESR-spectra can be well described by a single exchange narrowed Dyson line,\cite{Feher1955} i.e. a Lorentz line including some contribution of dispersion to the absorption due to non-zero electrical conductivity. Above $T_{\rm JT}$ the dispersion to absorption ratio even approaches values close to 1 (not shown) for the single-crystal measurements as the sample size is large compared to the skin depth.\cite{Ivanshin00} Because of the large linewidth $\Delta H$, which -- especially for X-band frequency -- is of comparable order of magnitude with the resonance field $H_{\rm res}$, the counter resonance at $-H_{\rm res}$ had to be included into the fit as described in Ref.~\onlinecite{Joshi2004}.

\begin{figure}[t]
\includegraphics[width=0.5\textwidth]{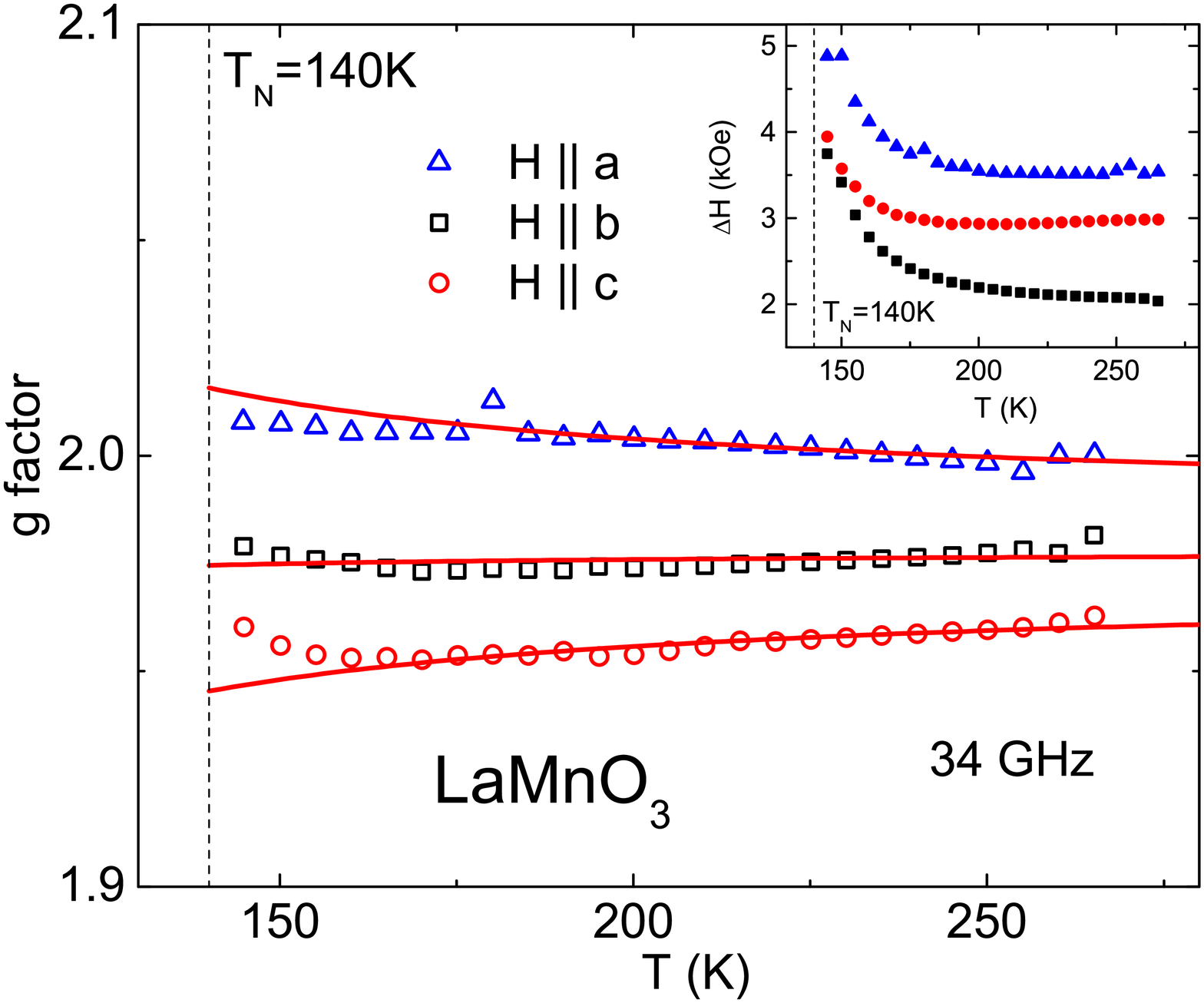}
\caption{\label{fig:gfactor} (Color online) Temperature dependence of the $g$ factor for different orientations measured at 34~GHz. Solid lines correspond to fit curves following Eq.~\ref{eqn:Eq1}. Inset: Temperature dependence of the ESR linewidth measured at 34~GHz.}
\end{figure}

The temperature dependence and anisotropy of the effective g-value obtained from the Q-band spectra is plotted in Fig.~\ref{fig:gfactor}. We want to emphasize that, although we are limited to temperatures $T<300$~K in the Q-band cryostat, an
evaluation of the effective $g$ factor has to be performed with data recorded at 34~Ghz, because at this frequency the resonance field $H_{\rm res}$ is larger than 12~kOe for the whole temperature range. Thus, it significantly exceeds the magnitude of the linewidth of the spectra (see inset of Fig.~\ref{fig:gfactor}), which is necessary to obtain reliable $g$ values.
As discussed previously, the $g$ tensor is determined by the local distortion of the MnO$_6$ octahedra, giving rise to a zero-field splitting of the $t_{2g}^{3} e_{g}^{1}$ (spin $S=2$) ground state of Mn$^{3+}$. To describe the data, we use the previously derived expressions \cite{Deisenhofer03,Kochelaev03,Ale03}
\begin{align} \label{eqn:Eq1}
\frac{g_{a,c}^{\rm eff}(T)}{g_{a,c}} &\approx 1+\frac{D}{T-T_{\rm
CW}}\left[(3\zeta-1)\pm
3(1+\zeta)\sin(2\gamma)\right]\nonumber \\
\frac{g_{b}^{\rm eff}(T)}{g_b} &\approx 1-\frac{2D}{T-T_{\rm
CW}}(3\zeta-1)
\end{align}
where $\zeta=E/D$ denotes the ratio of the rhombic and axial zero-field splitting parameters $E$ and $D$, respectively. The parameter $\gamma$ is given by the rotation angle of the MnO$_6$ octahedra in the $ac$-plane (crystallographic notation according to Ref. \onlinecite{Huang97}), and $T_{\rm CW}$ is the Curie-Weiss (CW) temperature.
We fix the angle $\gamma=13^\circ$ at the experimentally determined value\cite{Rodriguez98} and use $T_{\rm CW}$ = 58~K. As $T_{\rm CW}$ varies in literature\cite{Zhou99,Grando00} this value was determined by susceptibility measurements on the crystal used for the ESR measurements. The fit yields an axial parameter $D/k_{\rm B} =0.60(2)$~K and the $E/D$ ratio $\zeta=0.37(1)$ in agreement with the values found in La$_{0.95}$Sr$_{0.05}$MnO$_3$\cite{Deisenhofer03} and g-values $g_a = 1.988(1)$, $g_b = 1.978(1)$ and $g_c = 1.970(1)$, all slightly below 2 as expected for less than half-filled $d$ shells. \cite{Abragam1970}

\begin{figure}[b]
\includegraphics[width=0.5\textwidth]{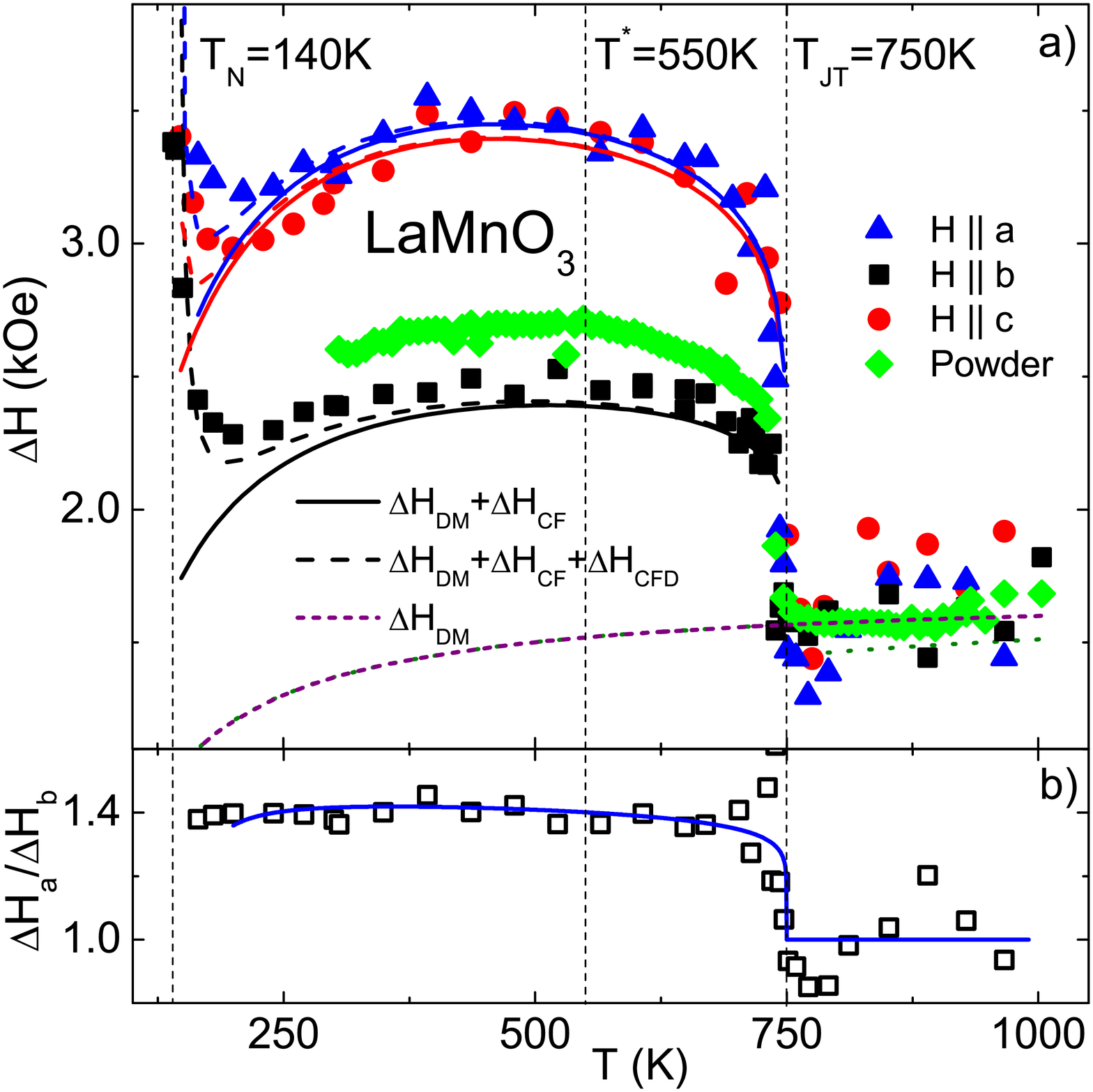}
\caption{\label{fig:dH} (Color online) a) ESR linewidth for different orientations of the single crystal and for the powder sample below and above $T_{\rm JT} =750 $~K. The lines correspond to fits following Eq.~\ref{eqn:Eq3}. b) Temperature dependence of the relative anisotropy ratio $\Delta H_a/\Delta H_b$. The solid line was obtained by simulating the relative anisotropy with Eq.~\ref{eqn:Eq3} and values obtained from fits to the linewidth.}
\end{figure}

The temperature dependence of the ESR linewidth $\Delta H$ in the range $100 \leq T \leq 1050 $~K is shown in the upper frame of Fig.~\ref{fig:dH} for X-band frequency and for comparison in the inset of Fig.~\ref{fig:gfactor} at Q-band frequency. In the low-temperature regime above $T_N = 140 $~K the ESR linewidth reveals a critical behavior on approaching $T_N$.  Within the orthorhombic $O^{\prime}$ phase the linewidth exhibits a strong anisotropy, which vanishes in a step-like manner at the  Jahn-Teller transition in agreement with the first-order character.
Like for the $g$ tensor, the anisotropy for $T < T_{\rm JT}$ mainly results from the zero-field splitting of the $S=2$ spin state of Mn$^{3+}$ due to spin-orbit coupling and crystal-electric field of the ligands.\cite{Kochelaev03,Deisenhofer03,Ale03}
The relative anisotropy ratio $\Delta H_a/\Delta H_b$ of the linewidth with the magnetic field parallel to the ferromagnetically coupled spins along the $a$-axis to the one along the antiferromagnetically coupled spins along the $b$-axis (see Fig.~\ref{fig:dH} b)) is rather constant at a value of 1.4 below the JT-transition, but approaches 1 as the system enters the paradistortive phase with only local distortions.\cite{Rodriguez98}

Again following Refs.~\onlinecite{Kochelaev03,Deisenhofer03,Ale03}, the anisotropy and temperature dependence of the linewidth is modeled by the expression:
\begin{equation} \label{eqn:Eq3}
\begin{split}
\Delta H (T,\theta,\phi) = \frac{\chi_0(T)}{\chi(T)}\{\Gamma_{\rm DM}
+ \left(\frac{T_{\rm JT}-T}{T_{\rm JT}}\right)^{2\beta}[\Gamma_{\rm CF} f_{\rm
reg}\\ + \Gamma_{\rm CFD}\left(\frac{T_{\rm N}}{6(T-T_{\rm N})}\right)^{\alpha}
f_{\rm div}]\}.
\end{split}
\end{equation}
Besides the crystal field (CF) parameters $\Gamma_{\rm CF}$ and $\Gamma_{\rm CFD}$ also the influence of the Dzyaloshinskii-Moriya (DM) interaction, $\Gamma_{\rm DM}$, has to be taken into account. Note that the DM interaction does not have any effect on the resonance field, because its first moment vanishes. The general temperature dependence is governed by the ratio of the single-ion susceptibility $\chi_0$ and the susceptibility $\chi$ of the interacting spins following Ref. \onlinecite{Huber99}. The regular CF contribution $\Gamma_{\rm CF}$ is empirically switched of at $T_{\rm JT}$ with a critical exponent $\beta$ which serves as a fit parameter. Moreover, a divergent CF contribution $\Gamma_{\rm CFD}$ has to be taken into account on approaching $T_N$ from above with a critical exponent $\alpha$. The angular dependences are given by $f_{\rm reg}$ and $f_{\rm div}$
\begin{eqnarray}
f_{{\rm reg}}(\theta ,\phi ) &=&A^{2}\left[1+\frac{3}{2}\sin ^{2}(\theta )\right]+f_{{\rm div}}(\theta ,\phi ) \nonumber \\
f_{{\rm div}}(\theta ,\phi ) &=&B^{2} +\frac{1}{2}\sin ^{2}(\theta )\left[4\gamma A B \cos (2\phi )-B^{2}\right].\nonumber
\end{eqnarray}
with the abbreviations $A=1+\zeta$ and $B=1-3\zeta$.
Here $\theta $ and $\phi $ are the polar and azimuth angles
between the external magnetic field and $z(b)$ and $x(c)$ axes
respectively.
Using the $E/D$ ratio derived from the temperature dependence of the g-factor, the temperature dependence of the linewidth can be described for all directions by Eq.~\ref{eqn:Eq3} with the temperatures $T_{\rm CW} = 58 $~K, $T_{\rm N} = 140$~K, and $T_{\rm JT} = 750$~K and the rotation angle in the $ac$-plane $\gamma=13^\circ$. The CF contribution is found to be $\Gamma_{\rm CF} (\infty) = 0.53(1)$~kOe with its critical exponent $\beta = 0.08(1)$. This contribution is comparable to the value found for La$_{0.95}$Sr$_{0.05}$MnO$_3$ ($\Gamma_{\rm CF} (\infty) = 0.57(2)$~kOe) but the critical exponent is only half of the exponent for the doped compound.\cite{Deisenhofer03} This reduced exponent supports a behavior close to a step-like transition of first-order character at $T_{\rm JT}$, whereas in the doped compound the smoother transition indicates rather a second-order character.
The DM contribution is found to be $\Gamma_{\rm DM} (\infty) = 1.70(1)$~kOe which is significantly higher compared to the contribution found in the doped compound ($\Gamma_{\rm DM} (\infty) = 1.0(1)$~kOe). The divergent CF contribution is found to be $\Gamma_{\rm CFD} (\infty) = 10(2)$~kOe with a critical exponent $\alpha = 1.3(2)$. The magnitude of this contribution is the same as in the doped sample, but the critical exponent -- $\alpha = 1.8(2)$ in La$_{0.95}$Sr$_{0.05}$MnO$_3$ -- is, in contrast to the doped sample, lower than the expected theoretical value of 1.5 indicating a shorter correlation range of the magnetic interactions.\cite{Kochelaev03}

\section{Discussion}

So far we have shown that the ESR data of untwinned LaMnO$_3$ single crystals are satisfactorily described in terms of crystal-field and Dzyaloshinskii-Moriya contributions in good agreement with previous evaluations of strontium-doped lanthanum manganite single crystals.\cite{Deisenhofer03,Ale03} The present results reveal, however, more details on the character of the transition from the $O^{\prime}$ into the $O$ phase, because it is not artificially broadened by inhomogeneities due to doping. For further discussion we refer to the measurements on LaMnO$_3$ powder samples which are also shown in Fig.~\ref{fig:dH}. The polycrystalline data match well those of the single-crystal, both concerning the transition temperature and the absolute value in the $O$-phase where the linewidth does not depend on the orientation of the magnetic field any more. Moreover, we recognize that the linewidth data in the $O$-phase coincide with the DM contribution determined independently in the $O^{\prime}$ phase. Here we recall that, due to the fact that the $g$-factor anisotropy already fixes the ratio $E/D$, the linewidth anisotropy yields basically the strength of the relaxation contributions $\Gamma_{\rm CF}$ and $\Gamma_{\rm DM}$ under the assumption that the DM contribution is approximately isotropic. This assumption is reasonable, because the Mn-O-Mn bond angles found along the $b$-axis as well as in the $ac$-plane are comparable,\cite{Huang97} so that any angular dependence of the DM interaction is averaged out.\cite{Deisenhofer03}

Such a continuation of the DM contribution from the $O^{\prime}$ phase into the $O$ phase is expected, because the tilting and buckling of the MnO$_6$ octahedra and the corresponding Mn-O-Mn bond angles, which determine the DM interaction, remain almost unaffected at the transition, although the distortion of the MnO$_6$ octahedra is strongly reduced above $T_{\rm JT}$.

However, recalling Eq.~\ref{eqn:Eq3} it is important to note that the dotted line $\Delta H_{\rm DM}$ in Fig.~\ref{fig:dH} has been calculated using a Curie-Weiss law for $\chi(T)$ with a Curie-Weiss temperature  $T_{\rm CW}=58$~K as measured from the susceptibility data in the $O^{\prime}$ phase. Indeed, the susceptibility significantly changes at $T_{\rm JT}$ to a higher Curie-Weiss temperature of about $T_{\rm CW} \approx 110$~K.\cite{Paraskevopoulos2000} Hence, the contribution $\Delta H_{\rm DM}$ is expected to drop by about 10\% at $T_{\rm JT}$ as indicated by the green dotted line. This suggests that even above $T_{\rm JT}$ the experimental linewidth still contains a non-zero crystal-field contribution due to residual distortions. Unfortunately, corresponding zero-field splitting parameters can not be obtained from a possible anisotropy within experimental uncertainty.

Residual distortions have been indicated by neutron-diffraction, as well. However, also the neutron data do not appear to be fully conclusive: on one hand one has to conclude that $D \propto c_{1}^{2} -c_{2}^{2}$ goes to zero, but $E \propto c_{1} c_{2}$ does not, as the difference between the orbital mixing coefficients $c_{1}$ and $c_{2}$ reduces upon heating and finally approaches zero at $T_{\rm JT}$.\cite{Rodriguez98, Wdowik11} On the other hand it is known that the CF parameters are very sensitive to the local distortion modes of the MnO$_6$ octahedra, i.e., $D\propto Q_{2}$, and $E\propto Q_{3}$. Keeping in mind this correlation, we can see from Fig.~5 of Wdowik \textit{et al.}\cite{Wdowik11} and Fig.~5 of Chatterji \textit{et al.}\cite{Chatterji03} that only $Q_{3}$ and, therefore, the parameter $E$ vanishes at $T \cong T_{JT} $, but $D$ does not. It is only strongly reduced above the phase transition, but remains at a finite small value. This apparent antagonism indicates that the behavior close to the phase transition at $T_{\rm JT}$ is more complicated and the applied models are oversimplified.

For a detailed analysis of the ESR relaxation behavior close to $T_{\rm JT}$ we remind that in the Kubo-Tomita approach, apart from any phase transition, the general temperature dependence of the ESR linewidth stems from the spin susceptibility following\cite{Kubo54,Huber99}
\begin{equation} \label{Dhkubotemp}
\Delta H_{\rm KT}(T) = \frac{\chi_0(T)}{\chi(T)} \cdot \Delta H_\infty,
\end{equation}
with the "reduced" high-temperature linewidth
\begin{equation} \label{Dhkubo}
\Delta H_\infty = \frac{1}{g \mu_{\textrm{B}}} \cdot \frac{M_2}{J},
\end{equation}
which is determined only by the ratio of the second moment
\begin{equation}\label{2Mom}
M_2 = \frac{\langle [ {\cal
H}_{\textrm{aniso}}, S^+] \, [S^-, {\cal H}_{\textrm{aniso}}] \rangle}
{\langle S^+ S^- \rangle}
\end{equation}
of the ESR spectrum due to any anisotropic interaction ${\cal H}_{\textrm{aniso}}$ and the isotropic exchange constant $J$. Thus, inserting the single-ion Curie susceptibility $\chi_0(T)\propto 1/T$ the product $\Delta H_{\rm KT}\cdot \chi\cdot T$ should be constant with temperature in case of temperature independent local interactions.
This means in case of rigid orbital order, the reduced linewidth is expected to be temperature independent except close to the divergence at $T_{\rm N}$.

\begin{figure}[t]
\includegraphics[width=0.5\textwidth]{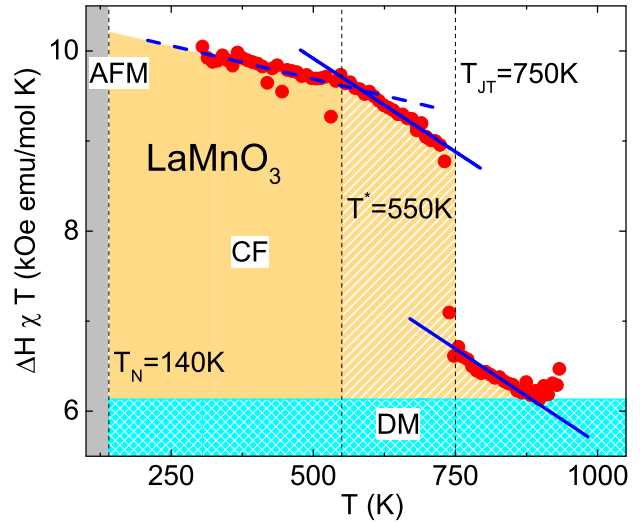}
\caption{\label{fig:Rdh} (Color online) Temperature dependence of the reduced ESR linewidth $\Delta H\chi T$ for the powdered sample. Lines are to guide the eyes, solid lines indicate equal slope. Light blue and orange areas show DM and CF contributions to the linewidth.}
\end{figure}

The reduced linewidth, shown in Fig.~\ref{fig:Rdh}, is calculated by the product of the linewidth measured in the powder sample, the susceptibility and the temperature. Although the reduced linewidth is not temperature independent, we can clearly distinguish the weakly temperature-dependent regime between room temperature and $T^*$ = 550~K, from that at higher temperature where the linear slope of the reduced linewidth becomes steeper, before the jump at the JT transition occurs. Interestingly, the slope for $T>T_{\rm JT}$ is the same as for $T^*<T<T_{\rm JT}$ up to about 900~K and levels off at higher temperatures, only. Indeed, the reduced linewidth decreases by about 10\% from 750~K to 900~K, which is just the excess contribution found from the extrapolation of the DM contribution from the $O^{\prime}$ phase into the $O$ phase. Thus, taking into account that the DM interaction mainly depends on the tilting angle and the Mn-O bond length, which is not significantly changing with temperature,\cite{Rodriguez98} we assign the remaining temperature dependent excess contribution of the linewidth to the zero-field splitting contribution which reflects the orbital occupation.

The temperature $T^*$ = 550~K coincides with the observation of a change at the ${}^{17}$O-site of the temperature dependence of the NMR Knight shift and the transverse relaxation rate (${}^{17} T_{2}^{-1} $) recently reported.\cite{Trokiner13} Moreover, a recent theoretical study estimated the orbital ordering temperature due to the Kugel-Khomskii superexchange mechanism as $T_{\rm KK} \approx  550$~K,\cite{Pavarini10} coinciding with $T^*$. To explain the experimental value of $T_{\rm JT} = 750$~K in their calculation these authors need to include the Jahn-Teller coupling.\cite{Pavarini10}
A quantitative analysis of the interaction between the Mn$^{3+}$-ions via the elastic field is given in the Appendix, where the interaction strengths via the elastic (phonon) field for ferro- and antiferrodistortive coupled Mn-ions lying in the $ac$-plane and along the $b$-axis are compared, respectively. As a result we find that orbital ordering in the $ac$-plane is more stable (by a factor of $2/\sqrt{3} $) than along the $b$-axis. This means that a temperature-induced melting of the orbital order is most likely to occur first along the $b$-axis. We therefore suggest from our ESR results that at $T^* = 550$~K the three-dimensional orbital ordering starts to melt, but the two-dimensional orbital ordering within the $ac$-planes remains intact up to  $T_{\rm JT} = 750$~K. Above 750~K the long-range orbital order is released, but short-range order fluctuations survive up to 900~K. This shows, that the interaction via the elastic field is one of the fundamental interactions defining the orbital order in LaMnO$_3$. This interaction has been missed in the previous discussion\cite{Pavarini10} and may improve the findings from theoretical studies in this system.
We believe that such a scenario of partial melting might also be realized in the layered dimer system Sr$_3$Cr$_2$O$_8$, where a cooperative JT distortion occurs already at 285~K, but strong lattice and orbital fluctuations persist down to about 120~K.\cite{Wang11,Wulferding11}

\section{Conclusions}
In summary, we investigated the spin relaxation dynamics of single- and polycrystalline LaMnO$_3$ up to high temperatures. The major contributions to the spin relaxation have been determined as $\Gamma_{\rm DM} (\infty) = 1.70(1)$~kOe from the Dzyaloshinskii-Moriya interaction and $\Gamma_{\rm CF} (\infty) = 0.53(1)$~kOe from the crystal-field contribution. By simultaneously evaluating the temperature dependence of the anisotropy of the g-value, the zero-field splitting parameters $D/k_{\rm B} =0.60(2)$~K and $E/D=0.37(1)$, could be verified for pure, detwinned LaMnO$_3$.
In the broad temperature regime $550 \leq T \leq T_{\rm JT}$ changes in the ESR linewidth have been observed and interpreted in terms of a partial melting of the orbital ordering along the $b$-axis at $T^*$ = 550~K due to the coupling of the orbital degrees of freedom with the elastic field. The orbital order within the $ac$-plane remains stable up to $T_{\rm JT}=750$~K, where an abrupt change into a fully isotropic behavior of the ESR linewidth is observed in agreement with the first-order type structural transition. Nevertheless, local order fluctuations are still present in the range $T_{\rm JT} \leq T \leq 900$~K. These findings show, that the interaction via the elastic field has to be taken into account in the discussion of the orbital order in LaMnO$_3$.

\begin{acknowledgments}
We thank Dana Vieweg for performing the SQUID measurements. This work is supported by the Deutsche Forschungsgemeinschaft (DFG) via the Transregional Collaborative Research Center TRR 80 (Augsburg, Munich, Stuttgart). MVE was funded by the subsidy allocated to Kazan Federal University for the state assignment in the sphere of scientific activities.
\end{acknowledgments}

\appendix

\section{Interaction via the elastic field}

The interaction between Mn$^{3+}$-ions $i$ and $j$ via the elastic field is given by the Hamiltonian \cite{Eremin1986}

\begin{widetext}
\begin{eqnarray}
{\cal H}^{ij} &=&\frac{(a+b)}{8\pi b(a+2b)r^{3} } \left[15\sigma _{\alpha \beta }^{i} \sigma _{\gamma \delta }^{j} n_{\alpha } n_{\beta } n_{\gamma } n_{\delta } -3(4\sigma _{\alpha \beta }^{i} \sigma _{\gamma \alpha }^{j}  +\sigma _{\alpha \alpha }^{i} \sigma _{\gamma \beta }^{j} +\sigma _{\gamma \beta }^{i} \sigma _{\alpha \alpha }^{j} ) n_{\beta } n_{\gamma }
+2\sigma _{\alpha \beta }^{i} \sigma _{\alpha \beta }^{j} +\sigma _{\alpha \alpha }^{i} \sigma _{\beta \beta }^{j} \right]  \nonumber \\
&+&\frac{1}{4\pi br^{3} } (3\sigma _{\alpha \beta }^{i} \sigma _{\gamma \alpha }^{j} n_{\beta } n_{\gamma } -\sigma _{\alpha \beta }^{i} \sigma _{\alpha \beta }^{j} )
\label{elastic_field}
\end{eqnarray}
\end{widetext}
Here $a=C_{11} $ and $b=C_{44} $ denote the elastic constants, the quantities $\sigma _{\alpha \beta } $ describe the interaction with the elastic tensor components $e_{\alpha \beta } $, which can be dynamic in general case and written via phonon operators. It is assumed that the electron-elastic field interaction is rewritten in the form ${\cal H}_{ee} =\sum \sigma _{\alpha \beta }  e_{\alpha \beta }$. In Eq.(~\ref{elastic_field}) $n_{\alpha}, n_{\beta}, n_{\gamma}, n_{\delta} $ are the components of the unit vector $\mathbf{r}$, connecting the Mn${}^{3+}$-ions. Pairs of indices $ij$ are omitted for clarity. The operators $\sigma _{\alpha \beta }$ are defined in the basis of orbital states ${\left| \vartheta  \right\rangle} $ and ${\left| \varepsilon  \right\rangle}$ as usual,\cite{Abragam1970} $V_{\rm e}$ is the electron-deformation parameter:
\begin{align} \label{Operatoren} \nonumber
\sigma _{xx} &=\frac{V_{\rm e} }{2} \left|\begin{array}{cc} {-1} & {\sqrt{3} } \\ {\sqrt{3} } & {1} \end{array}\right|,\sigma _{yy} =\frac{V_{\rm e} }{2} \left|\begin{array}{cc} {-1} & {-\sqrt{3} } \\ {-\sqrt{3} } & {1} \end{array}\right|, \\ \sigma _{zz} &=V_{\rm e} \left|\begin{array}{cc} {1} & {0} \\ {0} & {-1} \end{array}\right|
\end{align}

In our case the non-diagonal components $\sigma _{\alpha \beta }$ are absent and $\sigma _{xx} +\sigma _{yy} +\sigma _{zz} =0$; therefore Eq.(~\ref{elastic_field}) can be simplified for a pair of ions along the $z$-axis as follows:
\begin{align} \label{elastic_field_2} \nonumber
{\cal H}_{z-z}^{ij} &= \frac{(a+b)}{8\pi b(a+2b)r^{3} } \left[3\sigma _{zz}^{i} \sigma _{zz}^{j} +2\sigma _{\alpha \alpha }^{i} \sigma _{\alpha \alpha }^{j} \right]\\ &+\frac{1}{4\pi br^{3} } \left[3\sigma _{zz}^{i} \sigma _{zz}^{j} -\sigma _{\alpha \alpha }^{i} \sigma _{\alpha \alpha }^{j} \right]
\end{align}
The other two cases, i.e. along $x$- and $y$-axis, can be obtained from Eq.~\ref{elastic_field_2} by cyclic permutations of the indices.

As one can see from Eq.(~\ref{elastic_field_2}) this is indeed a strong interaction of antiferrodistortive nature. In particular, for nearest-neighbors $r \cong 4$~{\AA} -- using the values $V_{\rm e}=3$~eV, $a=5.7$~GPa and $b=6.9$~GPa as calculated in Ref.~\onlinecite{Nikiforov00} -- one obtains ${\cal H}_{z-z}^{ij} \cong -755$~cm$^{-1}$, i.e. $~1100$~K, which is of comparable to $T_{\rm JT} \cong 750$~K, and distortions of antiferrodistortive character within the $ac$-plane.\cite{Qiu05} Moreover, this estimate supports the idea that orbital order persists up to 1150~K in nanoclusters.\cite{Flesch2012}

Close to the orbital melting temperature $T\approx T_{JT}$ it is natural to expect that each MnO$_{6}$ fragment or complex migrates between three minima of the adiabatic potential. The corresponding three electronic states are given by:
\begin{equation} \label{el_states}
\begin{array}{l} {{\left| z \right\rangle} ={\left| 2z^{2} -x^{2} -y^{2}  \right\rangle} ={\left| \vartheta  \right\rangle} } \\ {{\left| x \right\rangle} ={\left| 2x^{2} -z^{2} -y^{2}  \right\rangle} =-\frac{1}{2} {\left| \vartheta  \right\rangle} +\frac{\sqrt{3} }{2} {\left| \varepsilon  \right\rangle} } \\ {{\left| y \right\rangle} ={\left| 2y^{2} -x^{2} -z^{2}  \right\rangle} =-\frac{1}{2} {\left| \vartheta  \right\rangle} -\frac{\sqrt{3} }{2} {\left| \varepsilon  \right\rangle} } \end{array}
\end{equation}
In general the electronic wave function at site $i$ is represented by:
\begin{equation} \label{wave_function}
{\left| i \right\rangle} =\cos \zeta _{i} {\left| z_{i}  \right\rangle} +\sin \zeta _{i} \cos \varphi _{i} {\left| x_{i}  \right\rangle} +\sin \zeta _{i} \sin \varphi _{i} {\left| y_{i}  \right\rangle}
\end{equation}
On the other hand  at $T<T_{\rm JT}$ the orbital order is usually described by the wave functions ${\left| i \right\rangle} =\cos \frac{\varphi _{i} }{2} {\left| \vartheta  \right\rangle} - \sin \frac{\varphi _{i} }{2} {\left| \varepsilon  \right\rangle} $ with  $\varphi _{i} \cong 109^{\circ} $.\cite{Abragam1970,Flesch2012}   

Using Equation~\ref{elastic_field_2} one gets
\begin{align} \label{elastic_field_3} \nonumber
\langle{\cal H}_{z-z}^{ij}\rangle &= \frac{V_{\rm e}^{2}}{8\pi b(a+2b)r_{ij}^{3}} \times \\ & \left[(15 +9 \frac{a}{b})\cos \varphi _{i} \cos \varphi _{j} - 2 \cos ( \varphi _{i} - \varphi _{j}) \right]
\end{align}
As one can see from Eq.~\ref{elastic_field_3} the energy of interaction via the elastic field attains a minimum at $\varphi_j = \varphi_i \pm \pi$  and $\varphi_i = 0, \pm \pi$ with energy $-755$~cm$^{-1}$. The negative sign indicates attraction. It is also clear that for the case $\varphi_j = \varphi_i$ we will get a higher energy whereby repulsion is also possible. This observation explains why in LaMnO$_3$ aniferrodistortive orbital order within the $ac$-plane accompanied by weak ferromagnetic spin order is more stable with respect to ferrodistortive orbital order along the  $b$-axis connected with antiferromagnetic spin order. As it was stressed in Ref.~\onlinecite{Flesch2012}, the orbital dependent superexchange mechanism alone cannot explain the high temperature of the orbital ordering transition in this compound.

On the other hand in a local coordinate system for octahedral fragments MnO$_{6}$ in the $ac$-plane the anharmonicity energies $g_{i} \cos 3\varphi_{i}$ and $g_{j} \cos 3\varphi_{j}$ have the minima at $\varphi _{i} =- 2\pi/3$ and $\varphi_{j} = 2\pi/3$, i.e. one gets alternating $\vartheta$-like orbital configurations with wave functions $|2x^2-z^2-y^2\rangle$  and $|2y^2-x^2-z^2\rangle$, or vice versa. The real value of about $\varphi _{i} \cong 109^{\circ} $ at $T_{N} < T < T_{JT}$ can be understood as a result of the competition of the local Jahn-Teller effect and the orbital-orbital interaction via the elastic (phonon) field. In mean-field approximation one can expect that the interaction via the phonon field modifies the profile of the adiabatic potential at each Mn site. Indeed, taking the values $g_i = -300$~cm$^{-1}$ (Ref.~\onlinecite{Nikiforov1980})
and $V_{\rm e}$, $a$, and $c$ like before \cite{Nikiforov00} we found that for a Mn-Mn pair in the $ac$-plane the minimum of the total energy
\begin{align} \label{elastic_field_4} \nonumber
 U_{z-z} &= \langle{\cal H}_{z-z}^{ij}\rangle + g_{i} \cos 3\varphi_{i} + g_{j} \cos 3\varphi_{j}
\end{align}
is shifted away from $\varphi _{i} =- 2\pi/3$ and $\varphi_{j} = 2\pi/3$ towards $\varphi_i = -\varphi_j = 113^{\circ}$ .
The corresponding energy coupling per one pair is $-485$~cm$^{-1}$, i.e. about 700~K.
Note that one can easily obtain the value $109^{\circ}$ by tuning the anharmonicity parameter $g_i$ or the electron-deformation parameter $V_{\rm e}$.

\end{document}